\documentclass[aps,prl,twocolumn,showpacs,superscriptaddress]{revtex4}

\usepackage {graphicx}
\usepackage{epsf}

\begin{document}

\title{Interatomic Coulombic Decay as a New Source of Low Energy Electrons in slow Ion-Dimer Collisions}

\author{W. Iskandar}
\email{iskandar@ganil.fr}
\affiliation{CIMAP, CEA - CNRS - ENSICAEN, BP 5133, F-14070, Caen cedex 5, France}
\author{J. Matsumoto}
\affiliation{Department of Chemistry, Tokyo Metropolitan University, 1-1 Minamiosawa, Hachiouji-shi, Tokyo 192-0397, Japan}
\author{A. Leredde}
\affiliation{Physics Division, Argonne National Laboratory, Argonne, Illinois 60439, USA}
\author{X. Fl\'echard}
\email{flechard@lpccaen.in2p3.fr}
\affiliation{LPC Caen, ENSICAEN, Universit\'e de Caen, CNRS/IN2P3, 14050 Caen Cedex 04, France}
\author{B. Gervais}
\affiliation{CIMAP, CEA - CNRS - ENSICAEN, BP 5133, F-14070, Caen cedex 5, France}
\author{S. Guillous}
\affiliation{CIMAP, CEA - CNRS - ENSICAEN, BP 5133, F-14070, Caen cedex 5, France}
\author{D. Hennecart}
\affiliation{CIMAP, CEA - CNRS - ENSICAEN, BP 5133, F-14070, Caen cedex 5, France}
\author{A.M\'ery}
\affiliation{CIMAP, CEA - CNRS - ENSICAEN, BP 5133, F-14070, Caen cedex 5, France}
\author{J. Rangama}
\affiliation{CIMAP, CEA - CNRS - ENSICAEN, BP 5133, F-14070, Caen cedex 5, France}
\author{C.L. Zhou}
\affiliation{CIMAP, CEA - CNRS - ENSICAEN, BP 5133, F-14070, Caen cedex 5, France}
\author{H. Shiromaru}
\affiliation{Department of Chemistry, Tokyo Metropolitan University, 1-1 Minamiosawa, Hachiouji-shi, Tokyo 192-0397, Japan}
\author{A. Cassimi}
\affiliation{CIMAP, CEA - CNRS - ENSICAEN, BP 5133, F-14070, Caen cedex 5, France}

\date{\today}

%abstract
\begin{abstract}
We provide the experimental evidence that the single electron capture process in slow collisions between O$^{3+}$ ions and neon dimer targets leads to an unexpected production of low-energy electrons. This production results from the interatomic Coulombic decay process, subsequent to inner shell single electron capture from one site of the neon dimer. Although pure one-electron capture from inner shell is expected to be negligible in the low collision energy regime investigated here, the electron production due to this process overtakes by one order of magnitude the emission of Auger electrons by the scattered projectiles after double-electron capture. This feature is specific to low charge states of the projectile: similar studies with Xe$^{20+}$ and Ar$^{9+}$ projectiles show no evidence of inner shell single-electron capture. The dependence of the process on the projectile charge state is interpreted using simple calculations based on the classical over the barrier model.
\end{abstract}


\maketitle
%introduction
When an atom or a molecule loses an electron from an inner-valence state, the excitation energy is usually not sufficient to remove a second electron. The resulting decay processes are thus limited to photon emission and molecular dissociation. However, if this inner-shell ionization occurs in the vicinity of another atom or molecule, it was theoretically predicted that a new decay channel called interatomic Coulombic decay (ICD) could open up \cite{Cederbaum97}. In this decay process, the energy resulting from a valence electron filling the inner-shell vacancy is transferred to the neighboring atom or molecule, where a low energy secondary electron is emitted. ICD was first observed in photoionization experiments for Ne clusters and dimers \cite{Marburger03, Jahnke04}, and soon after for a large variety of rare gas clusters (see \cite{Averbukh11, Hergenhahn11} and references therein) and for more complex systems such as water clusters \cite{Jahnke10, Mucke10}. 

%ICD& radiation dammage
As already demonstrated in \cite{Titze11}, ICD can also be induced by fast ion impact leading to inner-shell ionization. Being responsible for an additional production of low energy electrons (LEEs), it was pointed out that the ICD process could have a significant contribution in ion-induced radiation damage \cite{Kim11}. LEEs with energies bellow ionization thresholds induce bond cleavage in DNA bases, base-sugar, and sugar-phosphate units by dissociative electron attachment or dissociative electronically excited states production \cite{Boudaiffa00,Harbach13,Alizadeh13}. They can therefore be associated to an increased biological effectiveness of the ionizing radiation.
In fast collisions (with projectile velocities $v_p$ larger than the valence electrons velocities $v_e$) involving 0.1625 MeV/u He$^+$ projectiles, it has been shown that, due to ICD, the yield of secondary LEEs below 2 eV was increased by a factor 14 when using Ne$_2$ dimer targets as compared to Ne atomic targets \cite{Kim11}. Complementary experiments with 11.37 MeV/u S$^{14+}$, 0.150 MeV/u He$^{2+}$ and 0.125 MeV/u He$^+$ projectiles colliding on Ne$_2$ and Ar$_2$ dimer targets have confirmed the strong enhancement of LEEs production due to the ICD process in the fast collision regime \cite{Kim13}. In the present study we  provide the first experimental evidences that  the ICD process can also be a significant source of LEEs emission in low energy collisions ($v_p < v_e$), where inner shell single ionization of the target is expected to be negligible. 

%intro our work
We have investigated collisions with Ne$_2$ dimer targets using 2.81 keV/u O$^{3+}$, 3.37 keV/u Ar$^{9+}$ and 2.28 keV/u Xe$^{20+}$ projectiles. For the three collision systems, the projectile velocity was close to 0.35 a.u., well below the orbital velocity of the target active electrons. Capture of the target valence electrons by the projectile is thus expected to be the dominant process. The data analysis is focused on the Ne$^+(2p^{-1})+$Ne$^+(2p^{-1})$ fragmentation channel of the dimer target. As shown in previous studies for Ar$_2$ dimer targets in the same collision energy regime \cite{Matsumoto10,Iskandar14}, fragmentation in the Ne$^+(2p^{-1})+$Ne$^+(2p^{-1})$  channel can result from two competing double capture processes: from a 'two-site' double capture wherein one electron is removed from each atom of the dimer and leading directly to coulomb explosion (CE), and from a 'one-site' double capture populating Ne$_{2}^{2+}$ non dissociative molecular states. These transient states relax in a second step through radiative charge transfer (RCT) towards the same dissociative states as the 'two-site' double capture events. As detailed in \cite{Jahnke04, Santra00} the removal of a single electron from the $2s$ shell of a Ne atomic site will also lead to the Ne$^+(2p^{-1})+$Ne$^+(2p^{-1})$ dissociative channel through the ICD process. In the latter case, the transient Ne$_2^{+*}$($2s^{-1}$) molecular state decays by emitting an ICD electron from the neutral site of the excited dimer. As shown in Fig.1, the three processes referred here as CE, RCT and ICD end up in the Ne$^+(2p^{-1})+$Ne$^+(2p^{-1})$ fragmentation channel at different internuclear distances. They can therefore be distinguished by the different kinetic energy release (KER) in the fragmentation due to Coulomb repulsion between the two fragments.
%Figure1
\begin{figure}
\includegraphics[width=\columnwidth]{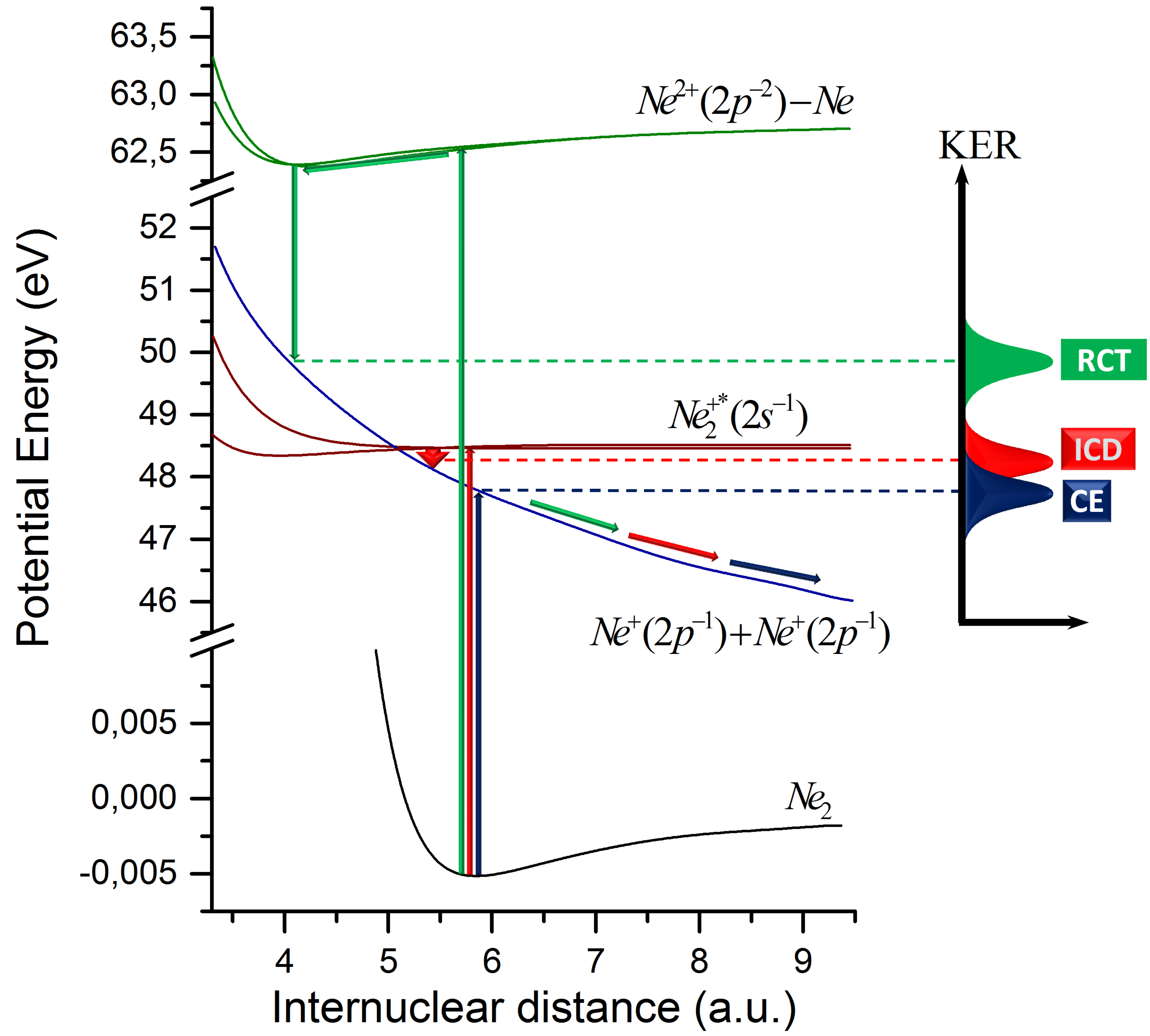}
\caption{(Color online) Illustration of the CE, RCT, and ICD processes in collisions between multi-charged ions and neon dimers. The  potential energy curves were adapted from \cite{Jahnke07, Stoychev08}. The KER distribution expected for each process is schematized on the right.}
\label{fig1}
\end{figure}

%Experimental setup
The three different projectiles O$^{3+}$, Ar$^{9+}$ and Xe$^{20+}$ ion beams were produced by an electron cyclotron resonance (ECR) ion source of the ARIBE facility, at GANIL, France. They were accelerated to 15 qkeV and guided towards the neon dimer target at the center of a COLTRIMS (cold target recoil ion momentum spectrometer) setup. The Ne$_2$ dimer target was provided by the supersonic expansion of neon gas at room temperature and 25 bar through a 30 $\mu$m nozzle. Recoil ions resulting from charge transfer were collected using the uniform electric field of the spectrometer and detected by a microchannel plate and delay-lines detector (MCPDLD) giving both the time and position of ion detection. Fragment ions from the dimers were identified by double-hit time of flight (TOF) coincidence measurements triggered by the detection of the scattered projectile on a second MCPDLD. The fragments momenta were calculated from the positions and TOF data by imposing momentum conservation restriction for optimal resolution and false coincidence events suppression\cite{Matsumoto11}. The kinetic energy release of the fragmentation was then inferred from the momenta of the fragment ions in the center-of-mass coordinates. The final charge state of the scattered projectiles was also determined using an electrostatic deflector combined with a measurement of their position on the second MCPDLD.

%Results
The data were first sorted as a function of the final charge state of the projectiles. Events involving two electrons transferred and kept by the scattered projectiles will be referred as true double capture (TDC), and events with only one electron kept by the projectiles as single-charge changing (SCC) events. TDC events can only arise from the CE ('two-site' double capture) and from the RCT ('one-site' double capture) processes. SCC events can correspond to single Auger electron emission by the scattered projectile following CE and RCT double capture processes, or to the single capture of a $2s$ electron followed by ICD. The KER distributions obtained for the Ne$^++$Ne$^+$ fragmentation channel in the three collision systems are displayed in Fig. 2. The KER spectra corresponding to TDC and SCC are on the left and right panels, respectively. 
%Figure2
\begin{figure}
\includegraphics[width=\columnwidth]{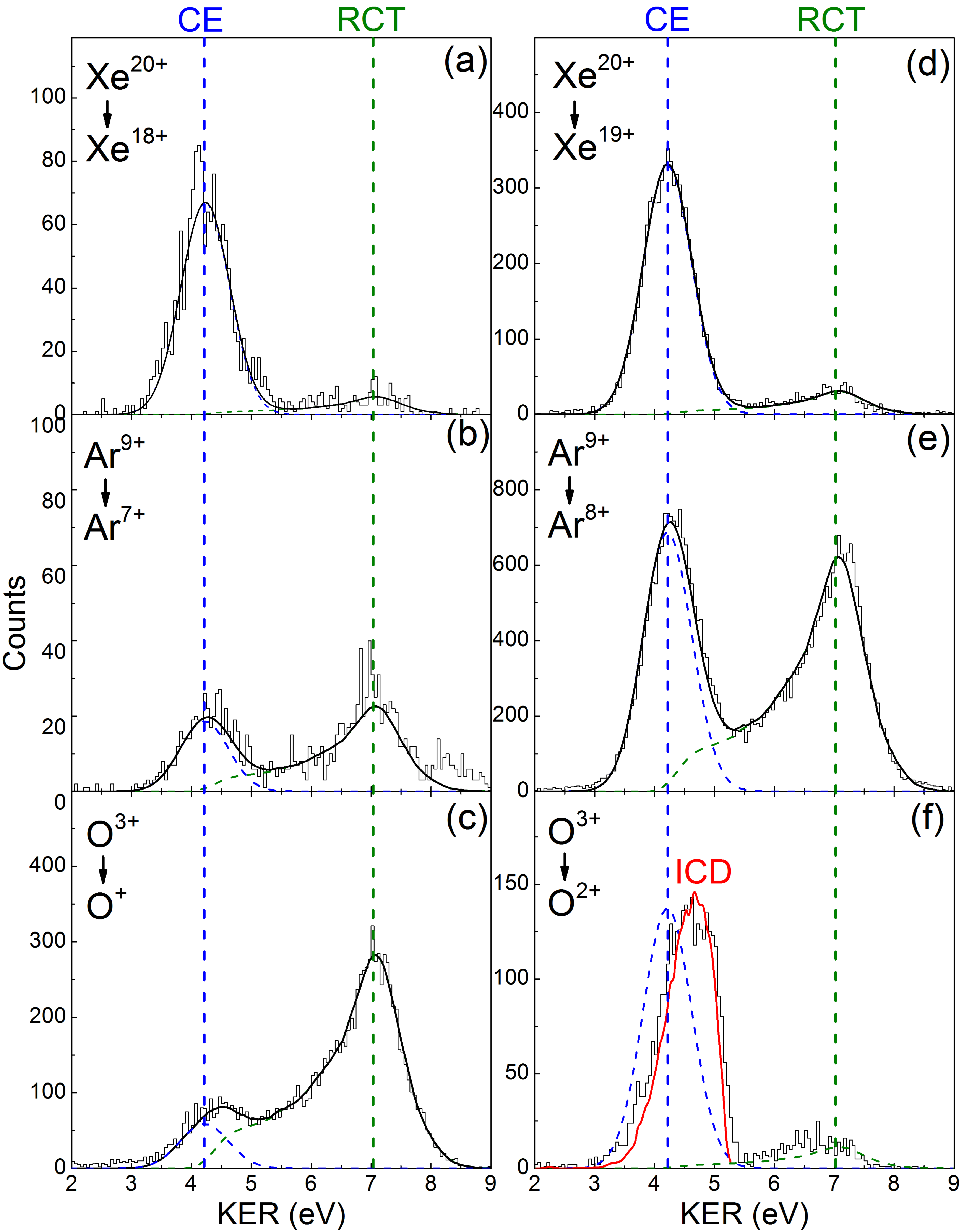}
\caption{(Color online) KER distributions obtained for the Ne$^+(2p^{-1})+$Ne$^+(2p^{-1})$ fragmentation channel with O$^{3+}$, Ar$^{9+}$ and Xe$^{20+}$ projectiles. Events associated to TDC and SCC are displayed on the left and right panels respectively. The dashed lines correspond to the fitted distributions for the CE and RCT processes and the solid lines to their sum. The red line displayed in the layer (f) shows the ICD distribution extracted from \cite{Jahnke04}. For each process, the peak position  is indicated by vertical lines.}
\label{fig2}
\end{figure}

%Peak identification
For a Franck-Condon transition at the internuclear distance R(Ne$_2$)=5.86 a.u., the expected KER is 4.6 eV when using a pure Coulomb approximation for the Ne$^+(2p^{-1})+$Ne$^+(2p^{-1})$ fragmentation channel. For the TDC events, the peak close to 4.2 eV could thus unambiguously be attributed to the direct CE process. Similarly, the RCT process, with transitions at the internuclear distance of the non-dissociative states R(Ne$_2^{2+}$) located between 4.05 a.u. and 4.35 a.u., results in a mean KER of 6.55 eV, close to the second peak of the KER spectra. For the ICD process, the available energy shared between the fragments and the emitted electron is equal to 5.3 eV. This corresponds to the difference between the ionization potential of a $2s$ electron (IP$_{2s}$ = 48.50 eV) and twice the ionization potential of a $2p$ electron (IP$_{2p}$ = 21.6 eV) from a neon atom \cite{NIST}.
The shape of the KER distribution associated to the CE process was approximated by a Gaussian curve whose mean and RMS values were obtained by fitting the KER spectrum of the Fig.2 (a), where the RCT contribution is almost negligible. The shape of the KER distribution associated to the RCT process was then deduced from the KER spectrum of the Fig.2 (c) by subtracting the CE contribution and smoothing the resulting histogram. The KER spectra of the Fig.2 corresponding to TDC events (layers (a), (b) and (c)), could be nicely adjusted with a combination of these CE and RCT distributions, displayed here with dashed lines. Similar adjustments of the KER spectra obtained in layers (d) and (e) for Xe$^{20+}$ and Ar$^{9+}$ show that these SCC events are also due to the CE and RCT double capture processes followed by Auger electron emission. The counts on the vertical scales, larger for the SCC than for the TDC spectra with both projectiles, indicate a strong preference for Auger versus radiative decay of the excited projectile after double electron capture.
The spectrum of the layer (f) obtained with O$^{3+}$ projectiles shows a very different result. Compared to the CE contribution, the first peak is clearly shifted towards higher KER values, where the ICD process is expected to arise. The shape of the KER distribution obtained for the ICD process in a previous experiment with the $2s$ shell photoionization of Ne$_2$ dimers \cite{Jahnke04} is displayed as a red line in the fig.2 (f), and fits almost perfectly the experimental data of this work. As previously shown in \cite{Jahnke04}, the shape of the KER spectrum indicates ICD electron energies ranging from 0 eV to 2 eV, with a maximum probability at 0.7 eV. Moreover, the ICD peak of the fig.2 (f) is larger, by about one order of magnitude (a factor of 8.5), than the RCT contribution leading to SCC. It shows clearly that, for this collision system, the emission of LEEs by the target resulting from the ICD process dominates by far the emission of Auger electrons from the projectile. This new observation of ICD for the O$^{3+}$-Ne$_2$ system demonstrates here that inner shell single electron capture followed by ICD can be one of the important processes at play in low energy collisions.

% Ratios RCT/CE
Beside the unexpected appearance of inner shell single-electron capture for the O$^{3+}$ projectiles, one can also clearly see the evolution of the relative contributions of the RCT and CE processes with the different charge states of the projectile. For higher charge states of the projectile, the CE process corresponding to a 'two-site' double capture dominates, while for lower charge states of the projectile, the RCT process associated to a 'one-site' double capture takes over. This behavior can be intuitively understood using simple geometrical considerations. The capture radii given by the classical over-the-barrier model (COBM) \cite{Niehaus86} for single and double electron capture from the $2p$ shell in collisions between Xe$^{20+}$ and atomic Ne targets are 12.54 a.u. and 9.73 a.u., respectively. This is in both cases larger than the internuclear distance R(Ne$_2$)=5.86 a.u. and should favor the removal of  electrons from both sites of the dimer. For O$^{3+}$ projectiles, the single and double electron capture radii are 5.63 a.u. and 4.58 a.u., respectively, what is comparable to the internuclear distance of the dimer. Electron capture from two different sites is then restrained to projectile trajectories close to the internuclear axis and 'one-site' double capture is then more probable than 'two-site' double capture. 
%Figure3
\begin{figure}
\includegraphics[width=\columnwidth]{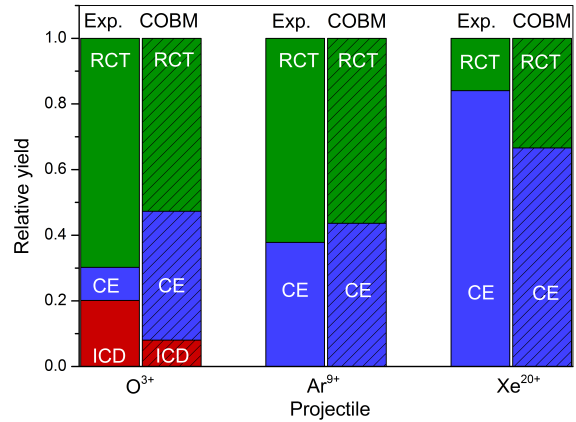}
\caption{(Color online) Experimental and calculated relative yields of the CE, RCT, and ICD processes contributing to the Ne$^+(2p^{-1})+$Ne$^+(2p^{-1})$ fragmentation channel. Both TDC and SCC processes are taken into account. Experimental uncertainties, not indicated here, were estimated to be about 0.05.}
\label{fig3}
\end{figure}

%calcul COBM
To investigate further the dependence of CE, RCT and ICD processes on the projectile charge state, we have performed simple calculations based on the COBM with the Ne$_2$ dimer target considered as two Ne atoms fixed in space. This method, that combines the COBM version of Niehaus \cite{Niehaus86} with Monte Carlo (MC) simulations, has already been described and successfully tested for Ar$^{9+}$ + Ar$_2$ low energy collisions \cite{Iskandar14}. In the calculations, the contributions of CE, RCT, and ICD simply arise from the two-site double capture probability, the one-site double capture probability, and the $2s$ shell single capture probability, respectively. For the RCT contribution, we also take into account the fact that 50\% of the one-site double capture events will statistically populate dissociative states leading to the Ne$^{2+}$ + Ne  fragmentation channel prior a possible RCT decay. The relative contributions of CE, RCT, and ICD given by the calculations for the different collision systems are compared to the experimental data in the Fig. 3. The predictions of our simple model are in quite good agreement with the experimental data. Calculations reproduce reasonably well the CE versus RCT contributions for the Ar$^{9+}$ and Xe$^{20+}$ projectiles, with a similar dependence of these contributions on the projectile charge state. Moreover, it is remarkable that, as for the experiment, the ICD process only appears for the O$^{3+}$ projectiles. The relative contribution of ICD predicted here by the model is 8.2\%, to be compared with the 20\% given by the experiment. Similarly, the calculated CE contribution for O$^{3+}$ projectiles is significantly overestimated. This discrepancy shows the limits of the COBM for low projectile charge states, where the hydrogenic approximation becomes too rudimentary for a precise estimate of capture probabilities. Still, these model calculations can help for a qualitative interpretation of the ICD yield with O$^{3+}$ projectiles.

%inner shell capture explanation
The sudden appearance of the inner shell single electron capture responsible for ICD can not be here simply explained by a geometrical approach. Capture radii for $2s$ electron capture from a neon atom by O$^{3+}$ and Xe$^{20+}$ projectiles are 2.12 a.u. and 4.02 a.u., respectively. With the capture radii given previously for a double electron capture, a basic calculation of geometrical cross sections leads to the quite similar ICD/RCT relative yields of 0.42 for O$^{3+}$ and 0.34 for Xe$^{20+}$. The explanation arises indeed when looking at the capture probabilities given by the COBM on the outgoing path (way out) of the collision. For Xe$^{20+}$ projectiles, the probability to capture the most bound $2p$ electron on the way out after the capture of a $2s$ electron is 84.1\%, and the probability to capture at least one of the $2p$ electrons is virtually 100\%. On the other hand, for O$^{3+}$ projectiles, the probability to capture the most bound $2p$ electron is only 11.5\%. In other words, $2s$ single capture by highly charged projectiles is systematically substituted by multiple capture. The difference comes from the low principal quantum numbers, $n$, populated on weakly charged projectiles ($n\sim 2$ for O$^{3+}$ projectiles) with only few free states available, compared to the high $n$ populated on highly charged projectiles ($n\sim 7-9$ for Xe$^{20+}$ projectiles). This is what makes inner shell single electron capture a significant process in low energy collisions for low charge states projectiles. At the same time, it also explains the weakness of Auger electron emission by the scattered projectile in collisions with O$^{3+}$ ions, while Auger decay is the dominant relaxation process for Xe$^{20+}$ and Ar$^{9+}$ projectiles. The population of only low $n$ quantum numbers by the two transferred electrons inhibits the Auger decay channel in the case of O$^{3+}$ projectiles. Finally, in order to have a larger view of the ICD contribution in low energy collisions with neon dimers, we have calculated with our model the expected yields of the ICD, RCT, and CE processes for O$^{q+}$ projectiles with q values between 2 and 5. The relative contributions of ICD to the  Ne$^+(2p^{-1})+$Ne$^+(2p^{-1})$ fragmentation channel were found to be $\sim$30\% for O$^{2+}$, $\sim$8\% for O$^{3+}$, $\sim$3\% for O$^{4+}$, and $\sim$1\% for O$^{5+}$, showing that the role of ICD increases when decreasing the projectile charge state.

In conclusion, we have shown that inner shell single electron capture followed by ICD can be a significant process in low energy collisions between multiply charged ions and neon dimers. For low charge states of the projectile, this process was found to have cross sections comparable to the ones associated with double capture processes. It is thus a new and unexpected source of LEEs in low energy collisions, that strongly overtakes Auger electron emission in the case of the  O$^{3+}$ + Ne$_2$ collision system that was investigated in this work. ICD is now recognized as a "universal" relaxation mechanism for atoms and molecules in weak interaction with a chemical environment. It is therefore expected that ICD, observed here for the first time in low energy ion collisions, should be responsible for an excess of LEEs emission in slow ion collisions with a large variety of targets such as rare gas dimers and clusters, or more complex systems such as water clusters. It may have significant implications in numerous phenomena involving low energy weakly charged ions, in particular at the end of the range of ions in matter.

We thank the CIMAP and GANIL staff for their technical support. This work was partly supported by TMU Research Program Grant.

\bibliography{biblio}

\end{document}